\documentclass[journal]{IEEEtran}
\usepackage{amsmath,amssymb,graphicx,cite}
\usepackage{xcolor}
\usepackage{hyperref}
\usepackage{multirow}
\begin{document}

\newcommand{\karmoose}{\color{red}}

\title{Discovering the Unequal Importance of Coded Bits in the Decoding of Polar Codes}

\author{Hossam Hassan\thanks{H. Hassan is with the Department of Electrical Engineering, Faculty of Engineering, Alexandria University, Alexandria, Egypt (email: \href{mailto:h.m.hassan@alexu.edu.eg}{h.m.hassan@alexu.edu.eg}).},
Ali Gaber\thanks{A. Gaber is with the Faculty of Engineering, Alamein International University (AIU), Egypt, and with the Department of Electrical Engineering, Faculty of Engineering, Alexandria University, Alexandria, Egypt (email: \href{mailto:aligaber@alexu.edu.eg}{aligaber@alexu.edu.eg}).},
Mohammed Karmoose\thanks{M. Karmoose is with the Wireless Intelligent Networks Center (WINC), School of Engineering and Applied Sciences (EAS), Nile University, Giza, Egypt, and with the Faculty of Engineering, Alexandria University, Alexandria, Egypt (email: \href{mailto:mkarmoose@alexu.edu.eg}{mkarmoose@alexu.edu.eg}).},
Noha Korany\thanks{N. Korany is with the Department of Electrical Engineering, Faculty of Engineering, Alexandria University, Alexandria, Egypt (email: \href{mailto:nokorany@hotmail}{nokorany@hotmail}).},}

\maketitle

\begin{abstract}
Polar codes are widely used in modern communication systems due to their capacity-achieving properties. This paper investigates the importance of coded bits in the decoding process of polar codes and aims to determine which bits contribute most to successful decoding. We investigate the problem via a brute-force search approach and surrogate optimization techniques to identify the most critical coded bits. We also demonstrate how mapping these important bits to the most reliable channels improves system performance with minimal additional cost. We show the performance of our proposed bit mapping in OFDM based systems, and demonstrate up to x7 gain in BER performance. 
\end{abstract}

\section{Introduction}
Polar codes, first introduced by Arıkan in 2009, are one of the popular Error-Correcting Codes (ECC) which are widely used in modern communication systems, where they theoretically achieve the capacity of Binary Input Discrete Memoryless Channels \cite{arkan2009channel}.  Polar codes enjoy the benefit of low-complexity encoding and decoding operations \cite{sarkis2016flexible}. Many low-complexity decoding algorithms exist for decoding polar codes, such as Successive Cancellation (SC) decoding, Successive Cancellation List (SCL) decoding \cite{tal2015list}, Automorphism Ensemble Decoding (AED) \cite{geiselhart2021automorphism}, Soft Cancellation (SCAN) decoding \cite{fayyaz2014low}, and Belief Propagation (BP) decoding \cite{arkan2008performance}\cite{qin2017polar}. Polar codes are currently the ECC code of choice for the control channels in 5G networks \cite{3gpp_212}.

Polar codes are initially designed on the premise that the system uses $N$ 
\textit{identical} channel, and polarization takes place across many such channels. In many systems, however, this assumption may not be true. For example, wireless communication experiences channel fading which can render the transmission over some time/frequency resource units more reliable than others. Assuming that the transmitter/receiver obtains an estimate of the channel, then we can no longer assume that the channel corresponding to $N$ different uses of the wireless channel are identical. In this work, we then ask the novel question: which of the coded bits in a polar-encoded codeword are the most important for the successful decoding of the codeword? In other words, assuming $N$ channels of different quality, which coded bits should be transmitted over the more reliable channels for better decoding performance?

We consider two approaches in order to formally answer this question. We assume a polar code with code rate ($K/N$), where $K$ and $N$ are the input block length and the codeword length of the polar code, respectively. In the first approach, we then ask the question: which $V$ of the $N$ coded bits are the most important $V$ bits for a successful decoding operation of the assumed polar decoder? More specifically, assume that the $N$ coded bits are communicated to the receiver via $N$ different channel realizations, where the system designer knows beforehand that a known set of $V$ out of those $N$ channels are more reliable than the remaining channels. With this knowledge, we attempt to know if there are $V$ bits out of the $N$ coded bits that are more important for the polar decoder, thus can be strategically transmitted through the more reliable $V$ channels to the receiver. 
As we show, this model resembles an Orthogonal Frequency Division Multiplexing (OFDM) communication system in a slow fading channel where the transmitter sends pilots to estimate the channel at $V$ subcarriers, and the receiver uses channel estimates from the $V$ pilots to estimate the channels at all $N$ resource units. Naturally, the estimation quality at the $V$ units will be higher than the remaining $N-V$ units, and thus are more reliable.

In the second approach, we ask the question: assuming $N$ channels with non-increasing reliabilities, which permutation of the $N$ coded bits is optimal to be used for mapping the $N$ coded bits over the $N$ channels? Such a permutation will determine which coded bits to be transmitted over which channel such that the decoder operation is optimized. An example of a practical system where such a model is relevant is again an OFDM communication system in a slow fading channel. However, in this case, the transmitter can send pilot signals at $N$ different subcarriers and then obtain channel estimation feedback from the receiver. The transmitter can then utilize the obtained channel knowledge to determine an ordering of the channels in terms of their reliabilities and can therefore map the $N$ coded bits to the channels using the aforementioned optimal permutation.   

Our work is different from the study of Unequal Error Protection (UEP) in polar codes. Namely, the UEP property specifies that data bits experience different levels of reliability in terms of decoding depending on their respective position in the encoding process\cite{hadi2018information}\cite{yao2024achievability}\cite{hanif2019efficient}. In contrast, our study is focused on coded bits, where we identify which coded bits are more important for the decoding operation of the polar code. Another related concept is the notion of bit reliability in a polar code. Specifically, input bits of the encoder may be ordered according to their reliabilities in the polarization process, which is then used to determine data bits and frozen bits in the encoding process. Computing bit reliabilities is difficult for large blocklengths, and therefore approximate solutions exist for determining reliability sequences such as \cite{xiong2020simplified} \cite{wu2022performance} \cite{trifonov2024design}. In contrast, our work attempts to identify which output bits of the encoder are more important to the decoding probability of the polar decoder.

\noindent{\bf Contribution}. Our first contribution is to determine the $V$ most important coded bits out of the $N$ output codeword for the polar decoder. Our work assumes a polar encoder based on the reliability sequence provided by the 3GPP standard for encoding the control channel of the 5G signal \cite{bioglio2020design} \cite{rosenqvist2019implementation} \cite{vermaimplementation}. We then consider a SC decoder and determine the most important $V$ coded bits. We first start by determining the most important bits for small values of $N$ using a brute-force search approach. To allow for larger values of $N$, we use surrogate optimization for finding a (potentially sub-optimal) set of $V$ important bits. We then showcase the usefulness of our findings in the context of a wireless communication system. We show that, by knowing the most important $V$ bits in a codeword, we can strategically map such bits in modulation symbols that are known to exhibit higher reliability in demodulation due to, e.g., having higher quality channel estimations. We show that such strategic mapping can enhance the performance of the system by a factor of up to $7$ at virtually no added cost.

Our second contribution is to determine a (sub-)optimal permutation of the $N$ coded bits which, when mapped to a set of $N$ communication channel with decreasing reliability, would lead to a better decoding performance of the polar code. We showcase the usefulness of this finding in the context of an OFDM-based wireless communication system under slow fading. In this case, both the transmitter and receiver obtain an estimate of the channel at all subcarriers can therefore order the subcarriers in terms of their reliabilities. Knowing the optimal permutation, the transmitter/receiver pair can now map the $N$ coded bits accordingly to the ordered set of subcarriers, enhancing the performance of the system by x4, again at virtually no added cost.

\section{System Model}

We consider a communication system that consists of a transmitter, a channel, and a receiver. 
%Two versions of the channel are considered: the Binary Symmetric Channel (BSC) and the Additive White Gaussian Noise (AWGN) channel.
At the transmitter, the input data bits $\boldsymbol{d}  \in \{0,1\}^k$ are first passed through a polar encoder which produces a coded bit sequence $\boldsymbol{c}  \in \{0,1\}^N$, 
% with input and output spaces
% \begin{equation}
%     \bold{d} \in \{0,1\}^k,\label{ali1}
% \end{equation}
% \begin{equation}
%     \bold{c} \in \{0,1\}^N,\label{ali2}
% \end{equation}
where $k$ is the number of information bits and $N$ is the number of coded bits. 
The encoding process is defined as:
\begin{equation}
    {\bf c} = {\bf u} \cdot {\bf G}_N,
\end{equation}
where $\bold{u} \in \{0,1\}^N$ is the input vector and ${\bf G}_N$ is the generator matrix built using the Kronecker product of the polarizing matrix \cite{arkan2009channel}. We assume $\mathcal{I} \subseteq [N]$ to be the set of positions used for information bits, selected based on the reliability sequence defined in the 3GPP standard \cite{3gpp_212}, while the rest of input bits are frozen to zeros. The input vector is therefore defined as
\begin{equation}
\begin{matrix}
     \bf{u}_\mathcal{I} = \bf{d}, & \bf{u}_{[N] \setminus \mathcal{I}} = \bf{0}.
\end{matrix}
\end{equation}
Coded bits are then processed by a mapper which permutes the coded bits according to a specific permutation. The mapper operation is described as 
\begin{equation}
\begin{matrix}
    \bar{c}_i = c_{\mathcal{P}(i)}, & i \in [N]
\end{matrix}  \label{eq::mapper}
\end{equation}
where $\mathcal{P} \subseteq [N]$ describes a suitable bit permutation.

The permuted bit sequence $\bar{\bf c}$ is passed through the channel which produces a received permuted bit sequence $\bar{\bar{\bf c}}$ along with a set of Log-Likelihood Ratio (LLR) values $\bf \ell$ for the permuted sequence. The channel operation is described below.
%signal is then transmitted over the selected channel.

At the receiver side, the received permuted bit sequence $\bar{\bar{\bf c}}$ and the LLR values $\bf \ell$ are passed through a demapper which performs the inverse operation of the mapper to restore the original order, i.e., 
\begin{equation} 
\begin{matrix}
    \hat{c}_{\mathcal{P}(i)} = \bar{\bar{c}}_i, &  \hat{\ell}_{\mathcal{P}(i)} = \ell_i & i \in [N]
\end{matrix}  \label{eq::demapper}
\end{equation}

After demapping, the restored bit sequence and LLR values are passed to a polar decoder, which we assume to be a Successive Cancellation (SC) decoder to recover the data bits.\\
\noindent \textbf{Channel.}  
In this paper, we consider two types of channels.
\noindent (a) \textbf{Binary Symmetric Channel (BSC).}  
In the BSC model, each transmitted bit can be flipped with a probability $p_i$. 
We refer to the $i$th channel is being \textit{more reliable} than the $j$th channel if $p_i < p_j$. Without loss of generality, we assume that $p_1 \leq p_2 \leq \cdots \leq p_N$.The flip probabilities are assumed to be known and may vary depending on the problem setup.

\noindent (b) \textbf{Additive White Gaussian Noise Channel (AWGN) and Slow Fading.} In this channel model, the permuted coded bit sequence $\bar{\bf c}$ is first modulated via Binary Phase Shift Keying (BPSK) modulation as follows 
\begin{equation}
    {\bf x} = 2 \bar{\bf c} - 1 \label{eq::modulation}
\end{equation}

The received signal over a block of $N$ symbols is given by
\begin{equation}
y_i = h_i x_i + n_i, \quad i \in [N] \label{eq::fading_equation}
\end{equation}
where the channel gains vector $\bf h$ is modeled as $\mathcal{N}(0, R_h)$ with $\mathbb{E} |h_i|^2 = \sigma_h^2$ and correlation matrix $R_h(i,j) = \mathbb{E} h_i h_j^\star$ that will be specified later on. The noise term $n_i$ is model as $\mathcal{N}(0, 1/\gamma)$, where $\gamma$ denotes the Signal-to-Noise Ratio (SNR). The output of the channel is demodulated to produce the coded bits $\bar{\bar{c}}$ based on Maximum Aposteriori (MAP) decoding.
\section{Determine Most Important $V$ Bits}
\begin{figure}
    \centering
    \includegraphics[width=\linewidth]{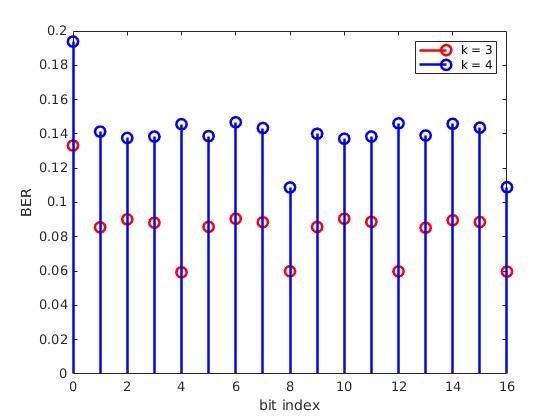}
    \caption{SC decoding performance (BER) versus index position for $N=16$, and $k=3,4$.}
    \label{fig:opt_N16_V1}
\end{figure}

The focus of this section is selecting the positions of the most important $V$ coded bits in the codeword. We start by assuming $N$ BSCs with the following transition probabilities
% is a key focus of this section.
% A channel is considered more important when it results in a lower bit error rate (BER) in the receiver, indicating that the transmitted bits are more likely to be received correctly. Conversely, a channel is considered less important if it has a higher BER, making the transmitted bits more prone to errors. In other words, the reliability of a channel determines whether it significantly affects the accuracy of the received codeword.

% This part is divided into three sections:

% \textbf{ Section 1: Optimization

% In this section, we assume a BSC with 
\begin{equation}
\boldsymbol{p}_i = 
\begin{cases}
    \boldsymbol{0} & \text{for } i = 1, \dots, V \\
    p & \text{for } i = V+1, \dots, N
\end{cases}
\end{equation}
where $p$ is a given transition probability. The goal is to find a mapping $\mathcal{P}^{1}$ that optimizes the decoding.
%in this setting.
Given the nature of the BSCs, we determine $\mathcal{P}^1$ by searching for the optimal selection of $V$ bits to be assigned to the first $V$ locations in $\mathcal{P}^1$ and then order the remaining $N-V$ in the remaining locations in $\mathcal{P}^1$.
% We perform the optimal selection step via surrogate optimization~\cite{queipo2005surrogate}. We first determine the optimal $V$ bits in a small polar code design and validate optimality of the selected bits by comparing to a brute-force search approach.
\noindent\textbf{Optimization.}
We consider a polar code with $N=16$ and we let $p=0.2$. We start by considering $V=1$, i.e., by setting $\mathcal{P}^1(1)$ to be equal to the chosen coded bit and assigning the remaining coded bits to the remaining $N-1$ locations in $\mathcal{P}^1$. Figure \ref{fig:opt_N16_V1} shows the Bit Error Rate (BER) of a SC decoder when choosing different indices to be assigned to $\mathcal{P}^1(1)$. By comparing the worst to best selection of a bit index, we notice a gain of $2.19$ and $1.7$ in the case of $k=3$ and $k=4$ respectively.
We then consider $V=2$. We examine the performance of all possible selections of index pairs to be
assigned to $\mathcal{P}^1(1)$ and $\mathcal{P}^1(2)$. Table \ref{tab:ber_results1} and \ref{tab:ber_results2} show the optimal selection of $\mathcal{P}^1$ for $V=1$ and $V=2$ in the case of $k=3$ and $k=4$, along with the gain in BER defined as the ratio between the maximum and minimum BER values obtained different $\mathcal{P}^1$.
We first observe that surrogate optimization provides $\mathcal{P}^1$ that performs close to the optimal selection. Interestingly, we also note that in the case of $k=3$, the optimal $V=1$ bit corresponds to $\mathcal{P}^1(1) = 4$, while the optimal $V=2$ bits correspond to $\mathcal{P}^1(1,2) = (6,12)$ which does not necessarily include the optimal bit for $V=1$. This implies that determining the optimal $\mathcal{P}^1$ for a given $V$ does not necessarily provide knowledge about the optimal $\mathcal{P}^1$ for other values of $V$. Finally, our simulations show that the optimal mapping leads to a reduction in BER values by a factor of $7$ with virtually no added communication or computation cost.

\begin{table}[t]
\centering
\caption{Optimal $\mathcal{P}^1$ Selection for $N=16$,$K=3$ and $V=1,2$ BSC with $p=0.2$}
\label{tab:ber_results1}
\begin{tabular}{|l|c|c|c|c|}
\hline
\textbf{Method} & \textbf{V} & \textbf{Optimal $\mathcal{P}^1$} & \textbf{BER} &\textbf{BER-gain} \\ \hline
Brute Force     & 1          & $\mathcal{P}^1(1) = 4$           & $0.0593$ & $(2.19×)$     \\ \hline
Brute Force     & 2          & $\mathcal{P}^1(1,2) = (6, 12)$   & $0.0187$ & $(5.83×)$     \\ \hline
Surrogate Opt.  & 1          & $\mathcal{P}^1(1) = 4$           & $0.0598$     &        \\ \hline
Surrogate Opt.  & 2          & $\mathcal{P}^1(1,2) = (14, 12)$  & $0.0202$      &       \\ \hline
\end{tabular}
\end{table}

\begin{table}[t]
\centering
\caption{Optimal $\mathcal{P}^1$ Selection for $N=16$,$K=4$ and $V=1,2$ BSC with $p=0.2$}
\label{tab:ber_results2}
\begin{tabular}{|l|c|c|c|c|}
\hline
\textbf{Method} & \textbf{V} & \textbf{Optimal $\mathcal{P}^1$} & \textbf{BER} &\textbf{BER-gain} \\ \hline
Brute Force     & 1          & $\mathcal{P}^1(1) = 8$ or $16$           & $0.1092$ & $(1.7×)$     \\ \hline
Brute Force     & 2          & $\mathcal{P}^1(1,2) = (8, 11),(12,15)$   & $0.0448$ & $(6.95×)$     \\ \hline
Surrogate Opt.  & 1          & $\mathcal{P}^1(1) = 16$           & $0.090$   &          \\ \hline
Surrogate Opt.  & 2          & $\mathcal{P}^1(1,2) = (8, 12)$  & $0.0431$   &          \\ \hline
\end{tabular}
\end{table}

Lastly, we determine a (sub-)optimal $\mathcal{P}^1$ using surrogate optimization for $N=128$, $k=24$ and $V=64$. We use this obtained permutation in assessing the system performance in a realistic OFDM-based communication setup as we see next.

\noindent\textbf{System Performance.} 
We consider an OFDM system with a slow fading where the channel correlation follows Jake's model, 
\begin{equation}
R_h(i,j) = J_0(2\pi f_D (i-j))
\end{equation}
where $J_0(x)$ is the zeroth-order Bessel function of the first kind and $f_D$ is the maximum Doppler frequency. We assume that channel estimation is performed by using $V$ pilots transmitted in locations $\mathcal{S} = \{1,\lfloor N/V \rfloor, 2\lfloor N/V \rfloor, \cdots, (V-1) \lfloor N/V \rfloor \}$. We then consider three different channel estimation cases: 1) perfect estimation, 2) Minimum Mean Square Error (MMSE) estimation and 3) linear interpolation.
\begin{figure}
    \centering
    \includegraphics[width=\linewidth]{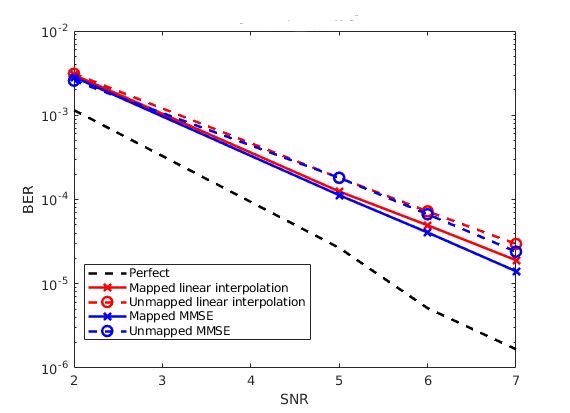}
    \caption{System performance comparison between optimal mapping and no mappring. $N=128$, $k=24$ and $V = 64$.}
    \label{fig:system_perf_1}
\end{figure}
Figure \ref{fig:system_perf_1} shows the performance of the system in terms of BER against SNR for a conventional system as well as a system with optimal mapping; the mapping used is the (sub-)optimal mapping obtained as described in the previous subsection. One can see that the system gains approximately up to 2x reduction in BER by employing the right mapping; this is obtained by virtually no added cost in terms of communication or computation effort.

\section{Determine an Optimal Permutation of the $N$ Bits}
In this section, our main focus is to minimize the BER by finding an optimal permutation of the coded bits denoted by $\mathcal{P}^2$. We start by finding a optimal permutation using optimization approachs. Then we assess the system performance using the optimum permutation in an OFDM based system.\\
 \noindent\textbf{Optimization}. We start by finding the optimal permutation, denoted $\mathcal{P}^{2}$, which results in the smallest BER when used to map the coded bits to symbols. In doing so, we assume
%Each permutation is evaluated by assuming transmission over 
$N$ Binary Symmetric Channels (BSCs), each with a different  flipping probability. We assume that the flipping probabilities are in decreasing order from $0.5$ to $0$ with a fixed step. 
%, equally spaced. 
Hence, the first symbol is transmitted through the least reliable channel (with flip probability $p_1 = 0.5$), while the last symbol is transmitted through the most reliable channel (with $p_N = 0$).

We consider the case of $N=128$ and attempt to find a (sub-)optimal permutation $\mathcal{P}^2$ using surrogate optimization. We notice that such optimization leads to a BER gain in the order of x4.

% For each permutation, the coded bits are mapped accordingly, transmitted through the BSCs, and then successive cancellation (SC) decoding is performed at the receiver. Finally, the resulting BER is computed. The permutation with the lowest BER is selected as the optimal permutation $\mathcal{P}^{2}$.

% figure here (waiting)

\noindent\textbf{System Performance.}
We assess the performance of the obtained optimal permutation $\mathcal{P}^{2}$ in an OFDM system with slow fading. 
% a more realistic communication scenario. The coded bits are modulated using BPSK and mapped according to $\mathcal{P}^{2}$, then transmitted through a fading channel with added Gaussian noise. The received signal is modeled as:
% 
% \begin{equation}
% y_i = h_i x_i + n_i, \quad \text{for } i = 1, \dots, N
% \end{equation}
% 
Here, we assume that the channel coefficients $h_i$
%is the fading coefficient of the $i$-th subchannel, and $n_i$ is additive white Gaussian noise (AWGN). The fading coefficients are assumed to be correlated according to Jake’s model. Furthermore, all channel coefficients 
are assumed to be perfectly known at the transmitter and receiver. Without loss of generality, the channels are sorted in increasing order of reliability such that
\[
|h_1|^2 \leq |h_2|^2 \leq \dots \leq |h_N|^2
\]
so that the $N$ channels in the OFDM system are in match with the order of the BSC channels used in the optimization step.
At the receiver, equalization is performed by dividing each received symbol by its corresponding channel coefficient to compensate for the fading effect. Subsequently, SC decoding is applied. The BER is then computed and compared to alternative mappings.

\begin{figure}
    \centering
    \includegraphics[width=\linewidth]{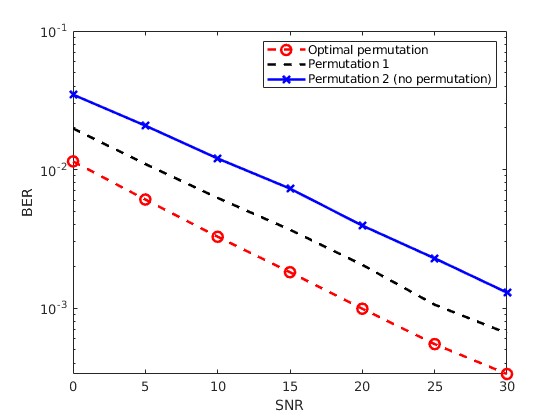}
    \caption{System performance comparison between three samples of  permutations. $N=128$, $k=24$.}
    \label{fig:enter-label}
\end{figure}

Figure\ref{fig:enter-label}shows how the BER changes with SNR for three cases: a good, a bad, and no permutation at all.
%using a good permutation, a bad permutation, and no permutation at all. 
The "Good permutation" line shows the best performance because it uses the optimized bit order found in the previous section. The "Bad permutation" line represents a random, less effective bit order. The "No permutation" line shows what happens when the bits are left in their original order. As the figure shows, using the right permutation improves the decoding performance—making the system much more reliable—without adding any extra complexity.

Our results demonstrate that using the optimal permutation $\mathcal{P}^{2}$ leads to significant enhancements in BER performance, often achieving these gains without increasing system complexity or communication overhead.
\section{Conclusion}
In this paper, we investigated how to improve the performance of polar decoding in wireless communication systems where channels have different levels of reliability. We showed that not all coded bits are equally important for decoding, and mapping the most important bits to the more reliable channels can help in terms of decoding polar codes and attain lower bit error rates.
%Our simulations showed that this approach could reduce the bit error rate by a significant factor (up to 7×) without adding extra complexity or cost. These findings can be directly applied to systems like OFDM in slow-fading environments. In future work, we plan to extend these ideas to longe’خايسخىسخى87+r polar codes and test them in more practical settings.}
%This paper presents a novel approach to identifying and leveraging the most important coded bits in polar codes. We propose brute-force and genetic algorithm-based methods to optimize bit placement for enhanced decoding performance. Future work includes extending the analysis to larger block lengths and real-world applications.
\bibliographystyle{ieeetr}
\bibliography{references}
\end{document}